\documentclass[twocolumn]{aastex631}

\usepackage{amsmath}
\usepackage{afterpage}
\begin{document}

\title{Probing Cosmic Isotropy with the FAST All Sky HI Survey}

\author{Yi-Wen Wu}
\affiliation{Institute for Frontiers in Astronomy and Astrophysics, Beijing Normal University, Beijing 100875, China}
\affiliation{School of Physics and Astronomy, Beijing Normal University, Beijing 100875, China; xiajq@bnu.edu.cn}

\author{Jun-Qing Xia} 
\affiliation{Institute for Frontiers in Astronomy and Astrophysics, Beijing Normal University, Beijing 100875, China}
\affiliation{School of Physics and Astronomy, Beijing Normal University, Beijing 100875, China; xiajq@bnu.edu.cn}



\begin{abstract}

This paper leverages the first released catalog from the FAST All Sky \textsc{Hi} Survey (FASHI) to examine the hypothesis of cosmic isotropy in the local Universe. Given the design of the overall FAST survey, the inhomogeneous detection sensitivity of FASHI is likely to introduce significant biases in the statistical properties of the catalog. To mitigate the potential influence of spurious clustering effects due to these sensitivity variations, we focus on extragalactic \textsc{Hi} sources within the sensitivity range of $[0.65, 1.0]$. This refined subsample is divided into ten distinct sky regions, for which we compute the two-point angular correlation functions (2PACF) over angular scales of $0.5^\circ < \theta < 10^\circ$. We apply the Markov chain Monte Carlo method to fit these 2PACFs with a power-law model and assess the statistical significance of the best-fit parameters for the ten FASHI sky regions by comparing them against results from mock catalogs generated under the assumptions of homogeneity and isotropy. Our findings indicate that the local Universe, as traced by the \textsc{Hi} sources in the FASHI survey, aligns with the cosmic isotropy hypothesis within a $2\sigma$ confidence level. We do not detect any statistically significant deviations from cosmic isotropy in the FASHI survey data.

\end{abstract}

\keywords{Observational cosmology (1146), Large-scale structure of the universe (902), Cosmic isotropy (320)}


\section{Introduction} \label{sec:intro}

The Cosmological Principle, which posits that the Universe is homogeneous and isotropic on large scales, serves as a foundational assumption in the standard cosmological model. This principle implies spatial uniformity in translation and rotation, making it a cornerstone of modern cosmology \citep{Peebles:1994xt}. 

With the development of advanced astronomical surveys, there has been growing interest in empirically testing the Cosmological Principle through various observational tracers \citep{Aluri:2022hzs}. These tracers include the cosmic microwave background \citep{Hansen:2004vq,Khan_2022}, galaxies \citep{Appleby:2014lra,deCarvalho:2021azj}, galaxy clusters \citep{Bengaly:2015xkw}, quasars \citep{Secrest:2020has}, supernovae \citep{Colin:2010ds,Deng:2018yhb}, X-ray galaxy clusters \citep{Migkas:2020fza,Ghosh:2016tbj}, gravitational waves \citep{Cai:2017aea}, galactic radios \citep{Bengaly:2017slg}, and future projects like the Chinese Space Station Telescope (CSST; \cite{Xu:2022gzj}).

Among these tracers, the neutral hydrogen (\textsc{Hi}) has played a particularly crucial role in astrophysical studies. Over the past two decades, significant surveys such as the Arecibo Legacy Fast ALFA survey (ALFALFA; \cite{Haynes_2011}) and the \textsc{Hi} Parkes All Sky Survey (HIPASS; \cite{Meyer:2004hr,Wong:2006pr}) have concentrated on \textsc{Hi} sources, offering valuable insights into the extragalactic population of the local Universe. Presently, the FAST All Sky \textsc{Hi} survey (FASHI; \cite{Zhang2023}) provides the most extensive extragalactic \textsc{Hi} catalog. Looking ahead, future projects like the Widefield ASKAP L-band Legacy All-sky Blind surveY (WALLABY; \cite{Koribalski:2020ngl}) are expected to become the most comprehensive \textsc{Hi} surveys, further enhancing our understanding of the local Universe.

\begin{figure*}[t]
\includegraphics[width=\textwidth]{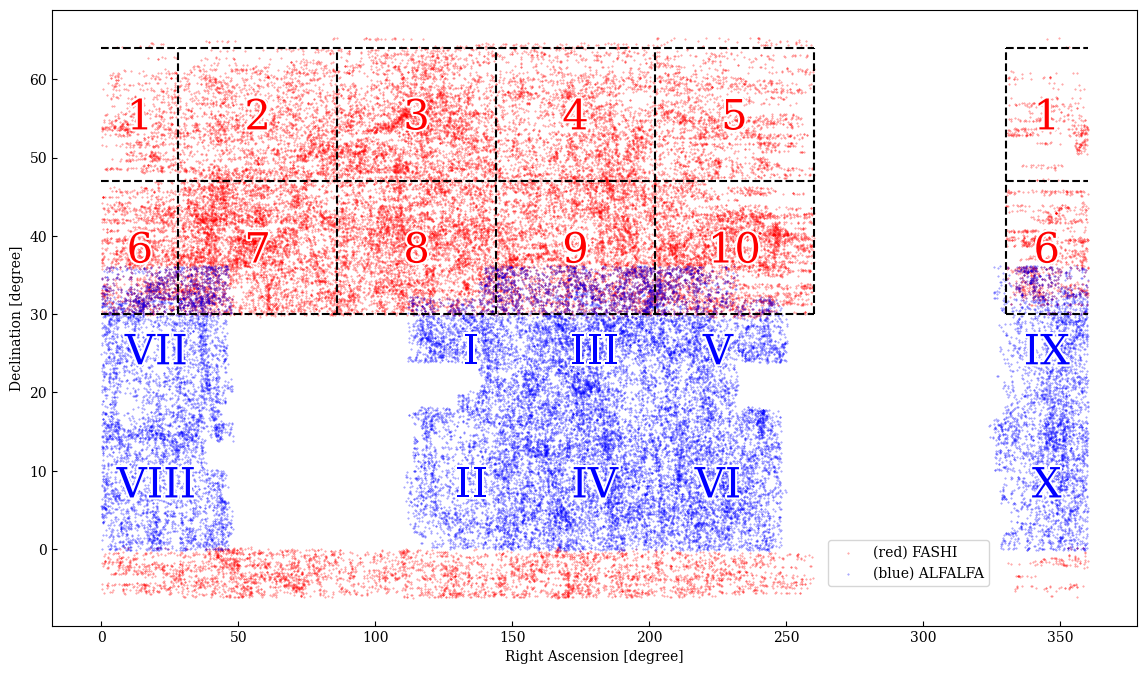}
\caption{The footprint of FASHI (red) and ALFALFA (blue) in Right Ascension and Declination is illustrated. FASHI strategically avoided most of the ALFALFA coverage to achieve a broader view of the Universe, with cross-matched signals primarily around DEC $\sim30^\circ$. The black dashed lines and Arabic numerals indicate the regions used to investigate isotropy in Section \ref{subsec:2PACF}, while the Roman numerals correspond to the regions outlined in \cite{Franco:2023rhd}.
\label{fig:map}}
\end{figure*}

In addition to testing the Cosmological Principle, \textsc{Hi} catalogs have been pivotal in studying the large-scale distribution of galaxies in cosmology \citep{Franco:2023rhd,Avila:2021dqv,Martin_2012}. The 21 cm emission line from \textsc{Hi} sources presents unique advantages that distinguish it from other tracers, due to its relatively simple underlying physics and the large number of sources it identifies, making it particularly valuable for cosmological investigations \citep{Zhang2023}. Furthermore, the bias parameter of \textsc{Hi} survey is close to 1.0 \citep{Martin_2012}, which enhances its reliability and accuracy as a cosmological tracer.

 As noted in \cite{Pan:2024xoj}, galaxy surveys face inherent limitations in efficiency and sky coverage. To address these challenges and enable more precise cosmological measurements, the 21 cm intensity mapping (IM) technique has been developed. This method leverages advanced radio telescopes, such as MeerKAT \citep{MeerKLASS:2017vgf}, Tianlai \citep{Chen:2012xu}, and the Canadian Hydrogen Intensity Mapping Experiment (CHIME, \cite{10.1117/12.2056962}), to probe the large-scale structure of the Universe. The integration of \textsc{Hi} galaxy surveys with 21 cm IM observations is expected to provide a more comprehensive understanding of cosmic evolution and structure formation.

Inspired by \citet{Franco:2023rhd}, this study utilizes the latest FASHI catalog to investigate the hypothesis of cosmic isotropy. We begin by selecting \textsc{Hi} sources within a well-defined sensitivity range. The survey area is then divided into ten distinct sky regions, where we calculate the two-point angular correlation function (2PACF) across various angular directions. By performing simulations under the assumption of a perfectly homogeneous and isotropic Universe, we compare the observed correlation functions with theoretical expectations, evaluating the consistency of the observed catalog with the Cosmological Principle.

The structure of this paper is as follows. Section \ref{sec:FASHI} provides an overview of the FASHI survey and outlines the methods employed in our analysis. In Section \ref{sec:2p}, we describe the procedure for calculating the 2PACF, detail the generation of mock maps replicating the characteristics of the FASHI survey, and present the corresponding results and analyses. Finally, Section \ref{sec:discu} offers a discussion of the findings and their implications.

\section{The Fast All Sky \textsc{Hi} Survey} \label{sec:FASHI}

The FAST All Sky \textsc{Hi} survey (FASHI) aims to cover the entire sky observable by the Five-hundred-meter Aperture Spherical radio Telescope (FAST), which is situated at a geographic latitude of $25^\circ39'$ and has a maximum observable zenith angle of $40^\circ$. The survey is anticipated to detect over 100,000 \textsc{Hi} sources across a declination range of $-14^\circ$ to $+66^\circ$, spanning approximately 22,000 square degrees, and reaching a redshift of about 0.35. FASHI delivers legacy data products with broad scientific applications, maximizing the potential for research across various domains. The survey is distinguished by its broader frequency coverage and greater detection sensitivity compared to most previous \textsc{Hi} surveys. Operating over a frequency range of $1.0 \sim 1.5$ GHz, it achieves a median detection sensitivity of approximately 0.76 mJy beam$^{-1}$ at a velocity resolution of 6.4 km s$^{-1}$ \citep{Jiang:2019rnj}.

FASHI samples a wide range of host galaxies in the local Universe, from very low \textsc{Hi}-mass dwarfs to gas-rich massive galaxies observed up to $z \sim 0.09$. The survey's first data release\footnote{https://fast.bao.ac.cn/cms/article/271/} includes 41,741 extragalactic \textsc{Hi} sources in more than 7,000 square degree with a high signal-to-noise ratio (S/N $> 6.5$) \citep{Zhang2023}. The detection rate is approximately 5.5 sources per square degree, slightly higher than that of the ALFALFA survey (around 3.15 deg$^{-2}$). The FASHI survey thus offers the largest extragalactic \textsc{Hi} catalog to date, providing an unbiased view of \textsc{Hi} content and large-scale structures in the Universe.

As this is the initial release, FASHI strategically avoided most of the regions covered by the ALFALFA survey, which spans ${\rm 0^\circ < RA < 260^\circ}$, ${\rm 330^\circ < RA < 360^\circ}$, ${\rm -6^\circ < DEC < 0^\circ}$, and ${\rm 30^\circ < DEC < 66^\circ}$. This strategy results in 3,620 well cross-matched sources at DEC $\sim 30^\circ$. The FASHI and ALFALFA surveys share consistent observational parameters, such as flux and velocity, but FASHI significantly outperforms ALFALFA in sensitivity, efficiency, and resolution. When combined, FASHI and ALFALFA offer a more comprehensive view of the local Universe, as illustrated in Figure \ref{fig:map}.

\begin{figure}[t]
    \centering
    \includegraphics[width=\linewidth]{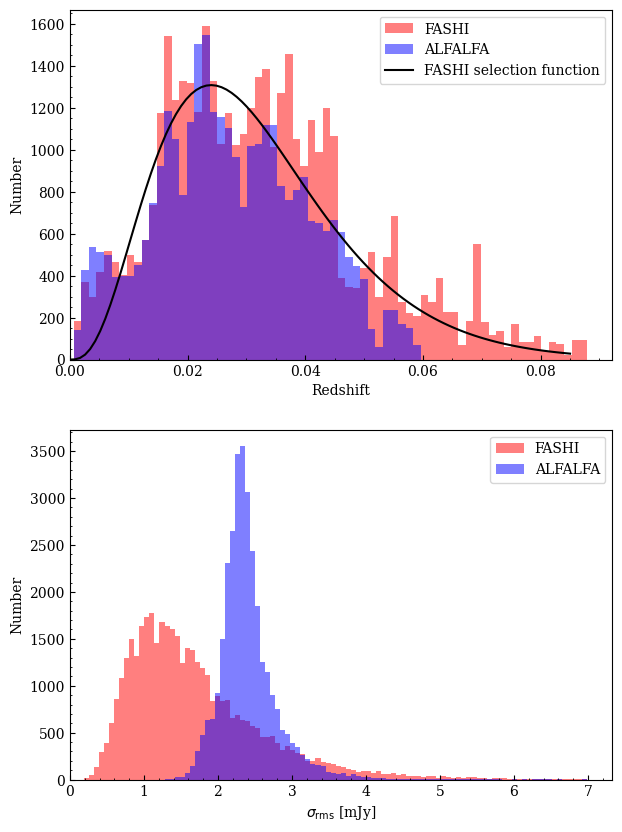}
    \caption{The histograms of redshift (upper panel) and measured spectral rms $\sigma_{\rm rms}$ (lower panel) for the FASHI and ALFALFA catalogs. The black solid line in the upper panel represents the selection function of FASHI catalog (Equation \ref{eq:selec}).}
    \label{fig:hist}
\end{figure}

In this study, we primarily focus on the extragalactic \textsc{Hi} sources within the declination range ${\rm 30^\circ<DEC<64^\circ}$ and at redshifts $z < 0.09$, resulting in approximately 36,879 \textsc{Hi} sources available for the 2PACF analysis (the first panel of Figure \ref{fig:hist}).

\subsection{Detection Sensitivity}

FASHI operates as a "schedule-filler" project, conducted during periods when no other programs are scheduled on FAST \citep{Jiang:2019rnj}. While this strategy effectively maximizes the observation time for \textsc{Hi} sample searches, it also introduces challenges in achieving uniform sky coverage.

In the first released catalog, we can use the measured spectral rms $\sigma_{\rm rms}$ and the source size parameters $ell_{\rm maj,\ min}$ to estimate the parameter {\rm RMS} which is related to the detection sensitivity of each source as follows:
\begin{equation}
    {\rm RMS} = \frac{\sigma_{\rm rms} \times 9.529287035639559}{\sqrt{(\pi ell_{\rm maj} \times ell_{\rm min})(\pi 2'.9^2) / 4}}~.
    \label{eq:sen}
\end{equation}
 These values are derived from the FASHI catalog \citep{Zhang2023}. The ellipse denotes the measurement aperture in right ascension (RA) and declination (DEC) space used to estimate the integrated flux of the source. It does not correspond to the intrinsic properties of the galaxy. The source size measurements were obtained directly from the FAST data, without applying deconvolution to account for the telescope's beam size.

From the FASHI catalog, the median detection sensitivity is approximately $0.76\ \text{mJy\ beam}^{-1}$, and the average is around $0.80\ \text{mJy\ beam}^{-1}$. However,  Figure 5 of \citet{Zhang2023} clearly illustrates that the detection sensitivity of \textsc{Hi} sources significantly varies across different regions due to the limitations of the adopted schedule-filler mode.

 It is important to note that the definition of sensitivity in \cite{Zhang2023} differs from the conventionally used definition. In their work, sensitivity is positively correlated with the root mean square (RMS), which quantifies the noise level. In contrast, we adopt the standard definition, where a higher RMS value corresponds to increased noise levels and lower detection sensitivity. Consequently, the uneven distribution of detection sensitivity directly influences whether a source can be included in the analysis, as we focus solely on high signal-to-noise sources. Although achieving a uniform sky coverage was not the primary goal for the first data release of FASHI, this uneven coverage can introduce artificial clustering features in large-scale structure analyses, posing challenges for cosmological studies.

We also present the distribution of the key parameter, the measured spectral rms $\sigma_{\rm rms}$, in the lower panel of Figure \ref{fig:hist}. The values of $\sigma_{\rm rms}$ span a wide range, from 0.2 mJy to 7 mJy, leading to significant variations in detection sensitivity across different regions. For comparison, we also include the $\sigma_{\rm rms}$ distribution for sources in the ALFALFA catalog (in blue). As shown, the ALFALFA distribution is much narrower, indicating more uniform detection sensitivity, although its median value is higher than that of the FASHI catalog. This suggests that while ALFALFA has slightly lower overall sensitivity, it achieves more consistent coverage across the survey.

Given these differences, it is crucial to account for the impact of detection sensitivity on clustering analyses. We therefore carefully select sources with appropriate detection sensitivity for performing cosmological statistical analyses, ensuring that the results are not skewed by the uneven sensitivity distribution.

\subsection{Angular Power Spectrum} \label{subsec:Cl}

To demonstrate the impact of detection sensitivity on clustering analysis, we calculate the angular power spectrum for subsamples selected based on varying detection sensitivity criteria. This approach allows us to assess how changes in sensitivity influence the clustering signal and ensures that our analysis remains robust against biases introduced by non-uniform detection sensitivity.

During the selection process, the following considerations are crucial: 1) The range of detection sensitivity should not be too broad, as significant variability in sensitivity can still introduce non-uniformity that may affect the clustering analysis of the catalog; 2) Conversely, the range should not be too narrow, as this would result in too few remaining sources, making the subsample unsuitable for robust statistical analysis; 3) We also ensure that the median sensitivity of the selected subsamples closely matches that of the original FASHI catalog to maintain consistency in the analysis. 

\begin{table}[t]
    \centering
    \caption{The number of sources in each sample is listed, along with the best-fit parameters for the bias parameter and shot noise, derived from the measurements of their angular power spectra. These results provide insights into how detection sensitivity impacts the clustering analysis and the reliability of the cosmological parameters inferred from the data.}
    \begin{tabular}{c|cccc}
        \hline
        Range & Number & $n$ [deg$^{-2}$] & $b_0$ & $C^{\rm SN}$ ($10^{-5})$\\
        \hline
        All & 36,879 & 5.87 &  $1.22 \pm 0.05$ & $5.70 \pm 0.11$\\
        $[0.55, 1.40]$ & 29,257 & 4.66 &  $1.20 \pm 0.05$ & $7.09 \pm 0.11$\\
        $[0.60, 1.10]$ & 24,229 & 3.85 &  $1.14 \pm 0.06$ & $8.35 \pm 0.17$\\
        $[0.65, 1.00]$ & 19,040 & 3.03 &  $1.02 \pm 0.06$ & $10.76 \pm 0.19$\\
        $[0.70, 0.93]$ & 13,339 & 2.12 &  $0.51 \pm 0.09$ & $14.09 \pm 0.27$\\
        \hline
    \end{tabular}
    \label{tab:ranges}
\end{table}

Finally, we consider four subsamples with varying ranges of detection sensitivity, as listed in Table \ref{tab:ranges}. As expected, when the sensitivity range is narrowed, the number of sources in each subsample decreases rapidly. Despite this reduction in source numbers, the redshift distributions of these subsamples remain nearly unchanged. This consistency suggests that the detection sensitivity is generally independent of redshift, making it possible to use these subsamples for reliable cosmological analysis without introducing redshift-dependent biases.

In our analysis of the angular power spectrum, we utilize the {\tt HEALPix} software \citep{Gorski:2004by} to generate source count maps with a resolution of $N_{\rm side} = 512$. To create the mask map, we downgrade the count map to a lower resolution of $(N_{\rm side} = 64$ and apply an apodization method. This method involves multiplying all pixels in the masked areas by a factor $f$ to mitigate edge effects and ensure smoother transitions at the boundaries. Finally, we use the public software {\tt NaMaster}\footnote{https://github.com/LSSTDESC/NaMaster} \citep{10.1093/mnras/stz093} to compute the observed angular power spectrum $C_\ell$. The pseudo power spectrum is measured up to $\ell_{\rm max}=450$, at these multipoles the magnitude of the estimated error bars is dominated by shot-noise. Due to the limitation of the Limber approximation, we set the minimal multipole $\ell_{\rm min}=20$. In these calculations, we carefully account for the coupling matrix $M_{\ell\ell'}$ that relates the true angular power spectrum $C_\ell$ to the pseudo angular power spectrum $\tilde{C}_\ell$, ensuring that the derived spectrum accurately reflects the underlying cosmological information.

We estimate the covariance matrix of the data points using the jackknife resampling method \citep{Scranton_2002}. This method involves dividing the data into $K$ patches, then creating $K$ subsamples by excluding each patch in turn. These patches are designed to cover roughly equal areas. Specifically, we list all the pixels covered by the survey and divide them into $K = 25$ patches, ensuring that while the patches may not have identical shapes, they do cover approximately equal areas (i.e., equal numbers of pixels).

The covariance estimator is expressed as:
\begin{equation}
Cov_{ij} = \frac{K-1}{K} \sum_{k=1}^{K} \left( C^k_{l_i} - \bar{C}_{l_i} \right) \left( C^k_{l_j} - \bar{C}_{l_j} \right)~,
\end{equation}
 where $C^k_{l_i}$ represents the angular power spectrum after excluding the $k$-th region, and $\bar{C}_{l_i}$ denotes the mean of all $k$ measurements at $l_i$. We also varied the number of patches $K$ to ensure the stability of the covariance matrix.

\begin{figure}[t]
    \centering
    \includegraphics[width=\linewidth]{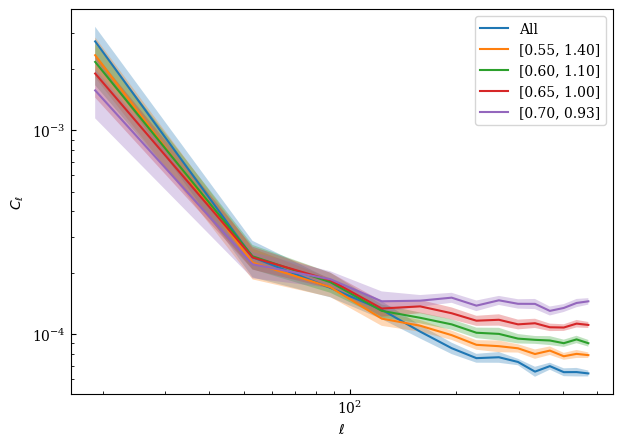}
    \caption{The obtained angular power spectra from subsamples with different ranges of detection sensitivity.}
    \label{fig:cls} 
\end{figure}

Figure \ref{fig:cls} displays the angular power spectra obtained from several subsamples, each evaluated with different ranges of detection sensitivity as listed in Table \ref{tab:ranges}. At smaller scales, narrowing the range of detection sensitivity leads to a rapid decrease in the number of sources within the subsample. This reduction significantly increases the shot noise, depicted as the elevated platform in the spectra. Conversely, at larger scales, which reflect clustering information, the angular power spectrum exhibits a different trend. When a narrower range of detection sensitivity is used, the slope of $C_\ell$ becomes more gradual, indicating a weakening of the clustering signal in the sample. This observation clearly demonstrates that the choice of detection sensitivity significantly influences the clustering signal, underscoring the importance of carefully selecting the sensitivity range in cosmological analyses.

\subsection{\textsc{Hi} Bias and Shot Noise}\label{subsec:bias}

To obtain quantitative results, we model the angular power spectrum using two free parameters: the constant bias $b_0$, which does not vary with redshift, and the shot noise $C^{\rm SN}$. We apply Markov chain Monte Carlo (MCMC) methods to fit the angular power spectrum using the following equation \citep{Xia:2009dr}:
\begin{equation}
    C_\ell^{\rm obse} = b_0^2 \times C_\ell^{\rm theo} + C^{\rm SN}~.
\end{equation}
Here, we derive the theoretical angular power spectrum $C_\ell^{\rm theo}$, which is defined as \citep{Dodelson:2003ft}:
\begin{equation}
    C_\ell = \frac{2}{\pi} \int k^2 P_m(k) \left[I_\ell(k)\right]^2 \, {\rm d}k~,
\end{equation}
where the window function is given by:
\begin{equation}
I_\ell(k) = \int_0^\infty b(z)n(z)j_\ell\left[k\chi(z)\right] \, {\rm d}z~.
\end{equation}
In this formulation, $P_m(k)$ is the matter power spectrum, and $b(z)$ represents the bias parameter with respect to the matter, which can be approximated as a linear constant $b_0 = b(z=0)$ at low redshifts. The term $j_\ell(k\chi)$ represents the spherical Bessel functions, while $\chi(z)$ denotes the comoving distance. The selection function $n(z)$ follows a FASHI-like distribution (depicted by the black solid line in the upper panel of Figure \ref{fig:hist}):
\begin{equation}\label{eq:selec}
n(z) \propto z^3 e^{-z/z_0}~,
\end{equation}
where $z_0 = 0.008$. It is important to note that this selection function is independent of the subsequent choices regarding detection sensitivity. 

By applying the Limber approximation \citep{Limber:1954zz}, the theoretical angular power spectrum at scales $\ell > 10$ can be expressed as
\begin{equation}
    C_\ell = \int \left[b(z)n(z)\right]^2\frac{H(z)}{c\chi^2(z)}P_m\left(k=\frac{\ell+1/2}{\chi},z\right){\rm d}z~.
\end{equation}
We computer $P_m$ and $C_\ell$ using the {\tt Colossus} numerical code \citep{2018ApJS..239...35D}, setting the cosmological parameters according to the best-fit values from the \textit{Planck} 2018 results \citep{Planck:2018vyg}. Finally, we use the publicly available code {\tt emcee} \citep{2013PASP..125..306F} to perform the MCMC analysis and obtain the best-fit values of those two free parameter.

Firstly, we analyze the entire FASHI catalog without applying any cuts on the detection sensitivity. As illustrated in Figure \ref{fig:cls}, the angular power spectrum $C_\ell$ for this full catalog exhibits a pronounced clustering signal at large scales. This results in a high value for the linear bias parameter: $b_0 = 1.22 \pm 0.05$ (68\% confidence level), indicating that these \textsc{Hi} sources are more clustered than the matter distribution, which deviates from previous expectations for \textsc{Hi} sources \citep{Marin:2009aw, Martin_2012}. We apply the same methodology to the ALFALFA catalog, which has a more uniform detection sensitivity. We obtain $b_0 = 0.98 \pm 0.05$ at 68\% confidence level, consistent with the results of \citet{Martin_2012}.

When applying a cut on detection sensitivity, the slope of the angular power spectrum becomes more gradual, as illustrated in Figure \ref{fig:cls}. The 68\% confidence level (C.L.) constraints on the linear bias parameter also show significant differences. This indicates that adjusting the detection sensitivity limits in the catalog effectively mitigates artificial clustering structures in the angular power spectrum. With a sensitivity range of $[0.65, 1.0]$, we obtain a linear bias of $b_0 = 1.02 \pm 0.06$ (68\% C.L.),  which aligns with the expected values for \textsc{Hi} sources \citep{Martin_2012}. When we further narrow the sensitivity range, the resulting linear bias is notably smaller than unity, suggesting that the clustering information is severely diminished. Therefore, we select the sensitivity range of $[0.65, 1.0]$ for the subsequent analyses to ensure accurate representation of the clustering properties.

On the other hand, using this subsample, we also obtain the best-fit value for the shot noise $C^{\rm noise} = 10.76 \times 10^{-5}$, which is consistent with the expected value of $\Delta \Omega / N$, where $ \Delta \Omega $ represents the surveyed area and $ N $ is the number of observed sources.

\begin{table}[t]
    \centering
    \caption{Details of the 10 regions.}
    \label{tab:10detail}
    \begin{tabular}{cccc}
        \hline
        & Area [deg$^2$] & source & density [deg$^{-2}$]\\
        \hline
        Full & 6,277 &  19,040 & 3.03\\
        Area 1 & 627 &  675 & 1.07\\
        Area 2 & 627 &  1,296 & 2.06\\
        Area 3 & 627 &  1,704 & 2.71\\
        Area 4 & 627 &  1,364 & 2.17\\
        Area 5 & 627 &  1,016 & 1.61\\
        Area 6 & 627 &  1,628 & 2.59\\
        Area 7 & 627 &  3,342 & 5.32\\
        Area 8 & 627 &  2,877 & 4.58\\
        Area 9 & 627 &  2,524 & 4.02\\
        Area 10 & 627 &  2,489 & 3.96\\
        \hline
    \end{tabular}
\end{table}

\section{Two-Point Angular Correlation Function} \label{sec:2p}

\subsection{2PACF Measurements}\label{subsec:2PACF}

After applying the detection sensitivity cut to the FASHI catalog, we are left with approximately $N_\textsc{Hi}=19,040$ \textsc{Hi} sources. These sources span an area of 6,277 deg$^2$ in the northern hemisphere, covering the sky within the declination range $30^\circ < \text{DEC} < 64^\circ$ and right ascension range $-30^\circ < \text{RA} < 260^\circ$. To investigate the isotropy of the local Universe, we divide the entire survey area into ten regions, each covering approximately 600 deg$^2$. These regions are illustrated in Figure \ref{fig:map}, and their geometric details are provided in Table \ref{tab:10detail}. We then analyze the clustering properties of these regions using the two-point angular correlation function (2PACF) to assess any anisotropies in the distribution of \textsc{Hi} sources.

Estimating the 2PACF among a set of footprints is a classical statistical problem widely used in astrophysical applications. Since the projection of the FASHI catalog onto the celestial sphere results in a 2-D map \citep{Giannantonio:2008zi}, we employ the public software {\tt PolSpice} \citep{Chon:2003gx} to calculate the edge- and noise-corrected 2PACF following the methodology outlined by \citet{Szapudi:1999za} and \citet{Szapudi:2001qj} in each region:
\begin{equation}
    \tilde{\xi}(\cos\theta) = \sum _{ij} f_{ij}(\Delta_i \Delta_j - N_{ij}),
    \label{eq:SS}
\end{equation}
where $\theta$ is the angle between two sky vectors, with $\cos\theta={\rm q}_1 \cdot {\rm q}_2$; $\Delta_i$ represents a set of bins that are linear in $\theta$, containing contributions from both signal and noise; $N_{ij}$ is the noise matrix. We define $f_{ij}=0$ unless the pair of pixels belongs to a specific bin in $\cos\theta$, ensuring $\sum_{ij} f_{ij}=1$. The range of $\theta$ is linearly divided into 19 bins, spanning from $0.5^\circ$ to $10^\circ$, to adequately capture the angular correlations at different scales.

To estimate the covariance matrix using the standard Monte Carlo (MC) simulation technique, we employ the public software {\tt FLASK} \citep{Xavier:2016elr} to generate 100 lognormal random catalogs under the homogeneity and isotropy hypotheses. These mock catalogs are designed to replicate the clustering features of the Local Universe. The input parameters required to generate these catalogs include: redshift $z = 0$, the linear bias $b_0 = 1.02$, the number of \textsc{Hi} galaxies $N_\textsc{Hi}$, and other best-fit cosmological parameters from the {\it Planck} Collaboration \citep{Planck:2018vyg}. These random catalogs are generated with the same number of sources and sky coverage as the real selected FASHI subsample. Additionally, they are divided into the same ten regions used in the analysis, allowing for a consistent comparison and covariance estimation across the different regions.

In each of the ten selected regions, we estimate the covariance matrix using the Monte Carlo (MC) method, defined as:
\begin{equation}
    Cov_{ij} = \frac{1}{M} \sum_{m=1}^{M} \left( \xi^m(\theta_i) - \bar{\xi}(\theta_i) \right) \left( \xi^m(\theta_j) - \bar{\xi}(\theta_j) \right)~,
\end{equation}
where $\xi^m(\theta_i)$ is the 2PACF measurement from the $m$-th random catalog at the angular separation $\theta_i$, and $\bar{\xi}(\theta_i) $ is the average 2PACF value across all $M=100$ random catalogs at the same angular separation. This approach allows us to quantify the uncertainty and correlations between different angular scales in the 2PACF measurements.

\begin{figure*}[ht!]
\centering
\includegraphics[width=\textwidth]{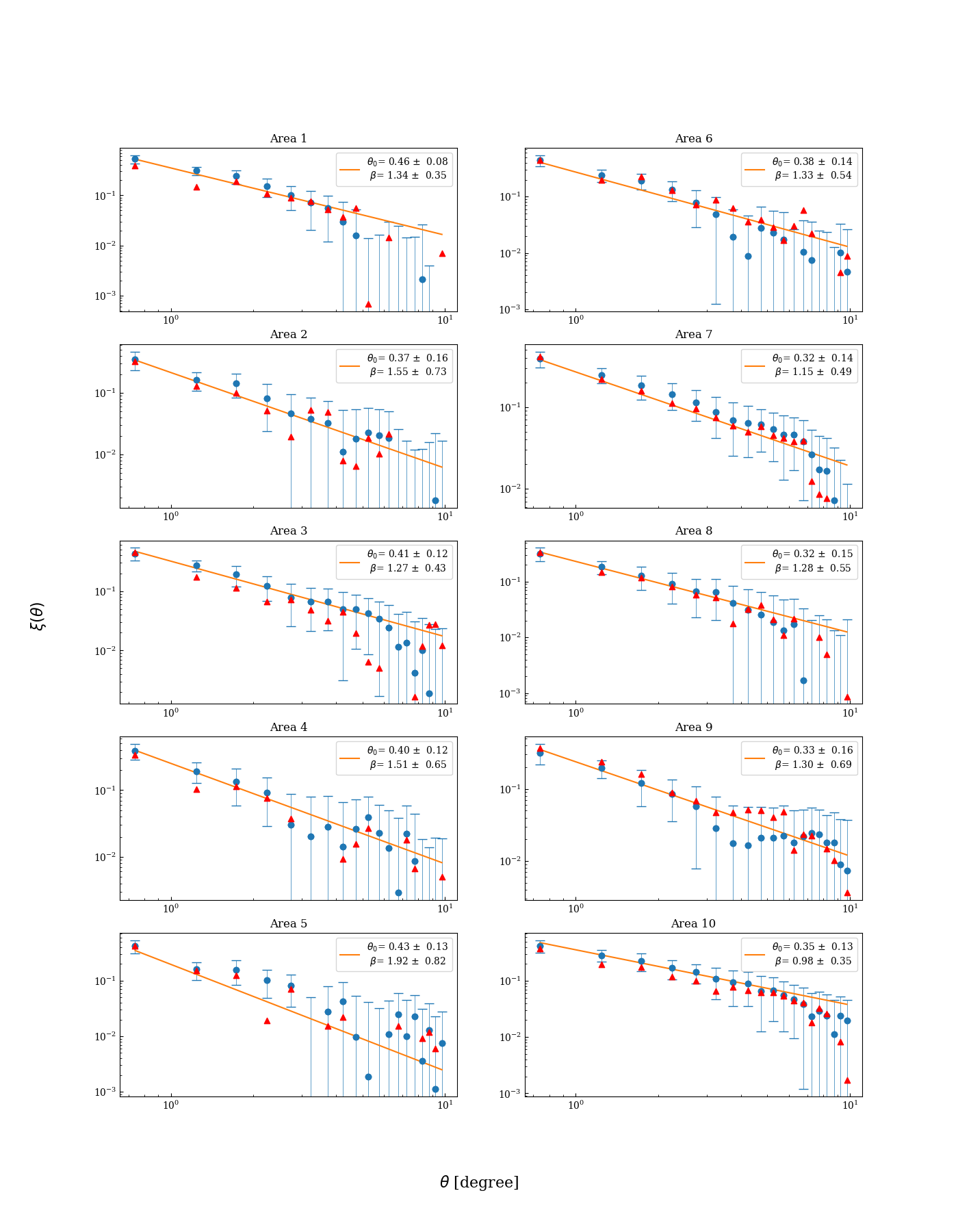}
\caption{We present the 2PACF measurements for the ten selected regions within the FASHI footprint. In these plots, the uncertainty is represented by the square root of the main diagonal of the covariance matrix, providing a visual indicator of the measurement precision.  The red triangle dots represent the 2PACF of the original FASHI catalog. Additionally, we overlay the theoretical 2PACF curves, which are derived using the best-fit values for the parameters $\theta_0$ and $\beta$.}
\label{fig:selec_2pacf}
\end{figure*}

\begin{table*}[ht!]
    \centering
    \caption{The 95\% C.L. constraint results on the parameters $\theta_0$ and $\beta$ from the 2PACF measurements in 10 regions. The original values are obtained directly from the original FASHI data, while the filtered values are from the sub-sample. In the last column, we also list the deviation of the fitted filtered $\beta$ from value from the expected value obtained from the random catalogs.}
    \label{tab:10result}
    \begin{tabular}{cccccc}
        \hline
         & original $\theta_0$ [degree] & original $\beta$ & filtered $\theta_0$ [degree] & filtered $\beta$ & $\beta$ std\\
        \hline
        Mock &  &  & $0.35\pm0.12$ & $1.40\pm0.76$ & \\
        Area 1 & $ 0.37 \pm 0.06 $ & $ 1.81 \pm 0.28 $ & $ 0.46 \pm 0.08 $ & $ 1.35 \pm 0.35 $ & -0.72$\sigma$\\
        Area 2 & $ 0.35 \pm 0.12 $ & $ 1.56 \pm 0.59 $ & $ 0.37 \pm 0.16 $ & $ 1.55 \pm 0.73 $ & 0.38$\sigma$\\
        Area 3 & $ 0.45 \pm 0.10 $ & $ 1.72 \pm 0.31 $ & $ 0.41 \pm 0.12 $ & $ 1.27 \pm 0.43 $ & -0.34$\sigma$\\
        Area 4 & $ 0.40 \pm 0.09 $ & $ 2.06 \pm 0.50 $ & $ 0.40 \pm 0.12 $ & $ 1.51 \pm 0.65 $ & 0.28$\sigma$\\
        Area 5 & $ 0.46 \pm 0.10 $ & $ 2.02 \pm 0.63 $ & $ 0.43 \pm 0.13 $ & $ 1.91 \pm 0.82 $ & 1.34$\sigma$\\
        Area 6 & $ 0.30 \pm 0.11 $ & $ 1.25 \pm 0.40 $ & $ 0.38 \pm 0.14 $ & $ 1.33 \pm 0.54 $ & -0.18$\sigma$\\
        Area 7 & $ 0.37 \pm 0.11 $ & $ 1.35 \pm 0.35 $ & $ 0.32 \pm 0.14 $ & $ 1.15 \pm 0.49 $ & -0.64$\sigma$\\
        Area 8 & $ 0.36 \pm 0.12 $ & $ 1.63 \pm 0.39 $ & $ 0.32 \pm 0.15 $ & $ 1.28 \pm 0.55 $ & -0.30$\sigma$\\
        Area 9 & $ 0.35 \pm 0.12 $ & $ 1.25 \pm 0.49 $ & $ 0.33 \pm 0.16 $ & $ 1.31 \pm 0.69 $ & -0.22$\sigma$\\
        Area 10 & $ 0.33 \pm 0.09 $ & $ 1.24 \pm 0.27 $ & $ 0.35 \pm 0.13 $ & $ 0.98 \pm 0.35 $ & -1.10$\sigma$\\
        \hline
    \end{tabular}
\end{table*}

In Figure \ref{fig:selec_2pacf}, we present the 2PACF measurements for each of the ten regions at angular separations between $0.5^\circ$ and $10^\circ$. The plots include the square root of the diagonal elements of the covariance matrix, $\Delta \xi(\theta_i) = \sqrt{Cov_{ii}}$, to indicate the uncertainty in the 2PACF measurements. We observe that: at large scales, the uncertainty in the 2PACF measurements is relatively high due to the limited sky coverage of the FASHI survey, which results in a larger noise contribution at these scales. Therefore, we do not consider the 2PACF measurements at scale $\theta > 10^\circ$. On the other hand, at small scales, the 2PACF measurements show stronger correlations among bins, and the measurements are more precise. This indicates that small-scale correlations are well-constrained despite the overall large uncertainties at larger scales. Given the strong correlations at small scales and the significant uncertainties at larger scales, we use the full covariance matrix in our likelihood analyses to account for all inter-bin correlations. This comprehensive approach will enhance the accuracy of the best-fit parameter evaluations and provide a more reliable assessment of the clustering features.

\subsection{2PACF Analyses}\label{subsec:result}

Unlike the previous analyses of the angular power spectrum, where theoretical predictions based on the standard cosmological model were employed, here we adopt a more straightforward approach. We fit the 2PACF measurements using a simple power-law form. This method has been validated in previous numerical studies and offers a direct way to characterize the clustering behavior observed in the data. The power-law form of the 2PACF is expressed as \citep{Peebles:1994xt}:
\begin{equation}
\xi(\theta) = \left(\frac{\theta}{\theta_0}\right)^{-\beta}~,
\label{eq:powerlaw}
\end{equation}
where $\theta_0$ is the angular scale at which the correlation function equals 1, marking the transition between linear and non-linear clustering regimes, and $\beta$ is the exponent that quantifies how the correlation decreases with increasing angular separation, with larger values of $\beta$ indicating a steeper decline and stronger clustering at small scales, resulting in a total of 1,000 regions for analysis. 

To quantify the angular clustering of \textsc{Hi} sources in the FASHI survey, we use the public software {\tt emcee} to perform a MCMC analysis. This analysis allows us to obtain constraints on the parameters $\theta_0$ and $\beta$ from the 2PACF measurements in the 10 selected regions of the FASHI footprint, which provide insight into how the clustering of \textsc{Hi} sources varies across different regions of the sky. 

Firstly, we use the random catalogs to estimate the 2PACF measurements and constrain the parameters \(\theta_0\) and \(\beta\). The resulting 68\% confidence level (C.L.) constraints are \(\theta_0 = 0.35 \pm 0.06\) and \(\beta = 1.40 \pm 0.38\). These values are consistent with those reported by \citet{Franco:2023rhd}, providing a reference standard for \(\theta_0\) and \(\beta\) under the assumption of a homogeneous and isotropic universe. This result serves as a benchmark for comparison: any significant deviation from these standard values in the actual data may suggest a breakdown of the isotropy hypothesis, indicating potential anisotropies or inhomogeneities in the spatial distribution of \textsc{Hi} sources. 

Next, we analyze the real 10 subsamples in the FASHI footprint. In Table \ref{tab:10result}, we summarize the 2PACF measurement results, presenting the 68\% confidence level (C.L.) constraints on the parameters \(\theta_0\) and \(\beta\) for each of the 10 regions. We also add the theoretical predictions of 2PACF $\xi$ using the best-fit values in each plot of Figure \ref{fig:selec_2pacf}. These analyses reveal that the constraints on \(\theta_0\) and \(\beta\) across these 10 regions are consistent with each other within a $2\sigma$ confidence level. This consistency suggests that the 10 FASHI regions are compatible with the hypothesis of statistical isotropy, indicating no significant deviation from the standard cosmological model in terms of angular clustering.

Specifically, the best-fit values of \(\theta_0\) across the 10 regions fluctuate between \(0.32^\circ\) and \(0.41^\circ\), with a median value of \(\theta_0 = 0.35^\circ\), which matches the reference value obtained from the random catalog. This consistency implies that the 2PACF measurements at angular separations \(0.5^\circ < \theta < 10^\circ\) primarily capture the linear clustering information, aligning well with the expected behavior of the large-scale structure in the Universe.

More importantly, a similar pattern is observed in the constraints on the parameter \(\beta\) across the 10 regions, which ranges from 0.98 to 1.91. The median value of \(\beta\) across these regions is approximately 1.36, nearly identical to the prediction from the random catalog. Additionally, we compare the \(\beta\) values derived from the FASHI data with those from the random catalogs, as listed in the last column of Table \ref{tab:10result}. This comparison is crucial for assessing the statistical significance of our findings and determining whether the observed clustering aligns with the expectations from random realizations. The results show that all 10 FASHI regions are compatible with the hypothesis of statistical isotropy within a \(2\sigma\) confidence level. We do not detect any statistically significant deviations from cosmic isotropy in the FASHI survey data.

 To assess the robustness of our findings, we performed the analysis using the original FASHI catalog. The 2PACF measurements for each region are presented in Figure \ref{fig:selec_2pacf}, which reveal larger fluctuations compared to those obtained from the filtered catalog. The derived values of $\theta_0$ and $\beta$ are summarized in Table \ref{tab:10result}. Notably, the value of $\beta$ indicates a stronger clustering amplitude relative to the filtered catalog, consistent with the bias parameter analysis discussed in Section \ref{subsec:bias}. Furthermore, the parameters derived from the original catalog exhibit larger deviations compared to both the filtered sample and the mock simulation, underscoring the critical role of sub-sample selection as highlighted in Section \ref{subsec:bias}.

Among the 10 regions analyzed, Area-5 and Area-10 stand out because the best-fit values of \(\beta\) in these areas deviate from the value obtained from the random catalog by more than others. This deviation suggests the potential presence of large-scale structures, such as super-voids or super-clusters, near these regions, which is worth to be verified in the future. Notably, Area-10 is situated close to Area-V of the ALFALFA survey (as shown in Figure \ref{fig:map}). To further investigate this anomaly, we calculated the best-fit value of \(\beta\) for Area-V in the ALFALFA survey and also found a lower value of \(\beta\), consistent with the findings reported by \citet{Franco:2023rhd}. This consistency strengthens the possibility of significant large-scale structures influencing the clustering signals in these regions.
\section{Discussion and Conclusion} \label{sec:discu}

The Cosmological Principle asserts that the Universe is isotropic on large angular scales, making it crucial to test this assumption through observations of large-scale structures. This study utilized data from the FASHI survey to investigate the validity of cosmic isotropy, providing an important assessment of the Cosmological Principle.

The first release of the FASHI catalog, which provides the most extensive \textsc{Hi} survey to date, holds significant potential for testing cosmic isotropy. However, the dataset poses challenges due to its survey strategy. In this study, we meticulously examined the detection sensitivity and selected an appropriate range to perform cosmological statistical analyses. As a result, we obtained a filtered catalog with number density and linear bias aligned with the ALFALFA survey. This subsample offers valuable prospects for future cosmological research using FASHI.
Here, we summarize our main conclusions in greater detail:

\begin{itemize}
    \item We calculate the angular power spectrum for four subsamples selected based on varying detection sensitivity criteria. This calculation learly demonstrates that the choice of detection sensitivity significantly influences the clustering signal, underscoring the importance of carefully selecting the sensitivity range in cosmological analyses. 
    \item We use two free parameters, the linear bias $b_0$ and the shot noise $C^{\rm SN}$, to fit the angular power spectrum. With a sensitivity range of $[0.65, 1.0]$, which includes 19,040 \textsc{Hi} sources and has a mean number density of approximately $3.02$ deg$^2$, we obtain the 68\% C.L. constraint on the linear bias of $b_0=1.02\pm0.06$, which aligns with the expected values for {\sc Hi} sources, while the best-fit value for the shot noise is consistent with the expected value.
    \item To assess the isotropy of the Universe, we compute the model-independent Two-Point Angular Correlation Function (2PACF) from the FASHI survey and compare it to theoretical predictions. The FASHI footprint is divided into 10 regions, each covering approximately 600 deg\(^2\), with an angular separation range of \(0^\circ < \theta < 10^\circ\), ensuring a suitable area for statistical analysis. We employ the Szapudi-Szalay estimator to calculate the 2PACF and analyze the angular correlation across ten different directions. 
    \item To quantify the clustering strength, we fit the 2PACF results to a power-law relation, characterized by the power index \(\beta\). The fitting process, performed using a Markov Chain Monte Carlo (MCMC) realization, yields the best-fit parameters, which are presented in Table \ref{tab:10result}. The median values obtained are around \(\theta_0 = 0.35^\circ\) and \(\beta = 1.36\), which are quite consistent with the predictions from the random catalogs under the homogeneity and isotropy hypotheses.
    \item All 10 regions of the FASHI survey are consistent with the hypothesis of statistical isotropy at the \(2\sigma\) confidence level. We do not observe any statistically significant deviations from cosmic isotropy in the FASHI data. However, in Area-10, the best-fit value of \(\beta\) deviates from the random catalog by more than \(1\sigma\), suggesting the possible existence of large-scale structures, such as super-voids or super-clusters, in this region. This deviation warrants further investigation to confirm the presence of such structures in future studies.
\end{itemize}

\section*{Acknowledgments}
We thank Ming Zhu for useful discussions about the FASHI catalog. This work is supported by the National Science Foundation of China under grant Nos. 12473004 and 12021003, the National Key R\&D Program of China No. 2020YFC2201603, the China Manned Space Program through its Space Application System, and the Fundamental Research Funds for the Central Universities.




%

\bibliography{sample631}{}
\bibliographystyle{aasjournal}



\end{document}